\documentclass[
aps, %
12pt,%
final,%
notitlepage,%
oneside,%
onecolumn,%
nobibnotes,%
nofootinbib,
superscriptaddress,%
noshowpacs,%
centertags]{revtex4}
\usepackage[dvips]{graphicx}
\usepackage[english]{babel}
\usepackage{amsmath}
\usepackage{amsthm} 
\usepackage{amssymb}	
\begin{document}
\title{Texture zeros in neutrino mass matrix}
\author{\firstname{Bartosz} \surname{Dziewit}}
\email[]{bartosz.dziewit@us.edu.pl}
\affiliation{Institute of Physics, University of Silesia\\
Uniwersytecka 4, 40-007 Katowice
}
\author{\firstname{Jacek} \surname{Holeczek}}
\email[]{jacek.holeczek@us.edu.pl}
\affiliation{Institute of Physics, University of Silesia\\
Uniwersytecka 4, 40-007 Katowice
}
\author{\firstname{Monika} \surname{Richter}}
\email[]{monikarichter18@gmail.com}
\affiliation{Institute of Physics, University of Silesia\\
Uniwersytecka 4, 40-007 Katowice
}
\author{\firstname{Sebastian} \surname{Zajac}}
\email[]{s.zajac@uksw.edu.pl}
\affiliation{Faculty of Mathematics and Natural Studies \\
Cardinal  Stefan Wyszynski University in Warsaw\\
Dewajtis 5, 01-815 Warszawa, Poland
}
\author{\firstname{Marek} \surname{Zralek}}
\email[]{marek.zralek@us.edu.pl}
\affiliation{Institute of Physics, University of Silesia\\
Uniwersytecka 4, 40-007 Katowice
}
\begin{abstract}
The Standard Model does not explain the hierarchy problem. Before the discovery of nonzero lepton mixing angle $\theta_{13}$ high hopes in  explanation of the shape of the  lepton mixing matrix were combined with  non abelian symmetries. Nowadays, assuming one Higgs doublet, it is unlikely  that this is still valid. Texture zeroes, that are combined with abelian symmetries, are intensively studied. The neutrino mass matrix is a natural way to study such  symmetries.
\end{abstract}

\maketitle
\section{Introduction}
In the Standard Model (SM) neutrinos interact with charged leptons. Interactions with the Higgs particle and interactions by neutral current with Z particle are not important for our purpose. In the flavour basis the Majorana neutrino Lagrangian effectively takes the form: 
\begin{equation}
\begin{split}
\mathcal{L}= & -\frac{e}{\sqrt{2}\sin \theta_W} \sum \limits_{\alpha=e,\mu,\tau} \bar{l}^\prime_{\alpha L}\gamma^{\mu} \nu_{\alpha L} W^-_\mu + h.c.  \\
& + 
\sum \limits_{\alpha,\beta = e, \mu, \tau} \left( M_{\alpha, \beta}^l \bar{l}^\prime_{\alpha R} l^\prime_{\beta L} + M_{\alpha, \beta}^{l^\ast} \bar{l}^\prime_{\beta L} l^\prime_{\alpha R} \right)+  \\
& +\frac{1}{2} \sum \limits_{\alpha, \beta = e, \mu, \tau} \left(
\mathcal{M}_{\alpha, \beta}^\nu \bar{\nu}^C_{\alpha R} \nu_{\beta L} + \mathcal{M}_{\alpha, \beta}^{\nu^\ast} \bar{\nu}_{\beta L} \nu^C_{\alpha R}
 \right).
\end{split}
\end{equation}
The neutrino mass matrix $\mathcal{M}^\nu$ is an arbitrary symmetric three dimensional complex matrix. It  is diagonalized by the orthogonal transformation with a complex matrix:
\begin{equation}
\left( U^{\nu \dag} \mathcal{M}^\nu U^{\nu\ast} \right)_{i j} = \delta_{i j} \, m^{\nu}_{i}  , \quad\quad m^{\nu}_{i} \geq 0. \label{macierzmasowa}
\end{equation}
The charged leptons mass matrix $M^l$ is an arbitrary three dimensional complex matrix diagonalized by the biunitary  transformation:
\begin{equation}
\left( U^{l\dagger}_L M^l U_R^l \right)_{\alpha \beta} = \delta_{\alpha \beta} m^l_{\alpha}, \quad \quad m^l_{\alpha} >0.
\end{equation}
As usually, let us introduce the physical basis in which the mass matrices become diagonal: \begin{align}
    \nu_{\alpha L} &= \sum \limits_{i=1,2,3} U^\nu_{\alpha i} \nu_{i L}  &\Rightarrow &  &\nu_{\alpha R} & = \sum \limits_{i=1,2,3} U^{\ast \nu}_{\alpha i} \nu_{i R}, 
    \intertext{and also:}
    l^\prime_{\alpha L} &= \sum \limits_{\beta=e,\mu,\tau} \left( U^l_L \right)_{\alpha \beta} l_{\beta L}  &\textrm{and}&  &l_{\alpha R}^\prime& = \sum \limits_{\beta=e,\mu,\tau} \left( U^l_R \right)_{\alpha \beta} l_{\beta R}.
\end{align}
After such transformation, the Pontecorvo-Maki-Nakagawa-Sakata mixing matrix appears in the charged current interactions:
\begin{equation}
\mathcal{L}_{CC}=-\frac{e}{\sqrt{2}\sin \theta_W} 
\sum_{\substack{\alpha=e,\mu,\tau \\ i=1,2,3} }\bar{l}_{\alpha L}\gamma^{\mu} \left(U_{PMNS}\right)_{\alpha i} \nu_{iL} W^-_\mu + h.c, \label{l300}
\end{equation}
\begin{equation}
\left(U_{PMNS}\right)_{\alpha i}= \sum \limits_{\beta=e,\mu,\tau} \left(U_L^{l } \right)^\dag_{\alpha \beta} \left(U^\nu \right)_{\beta i}, \label{macierzU}
\end{equation}
One should mention that the mixing matrix \eqref{macierzU} is expressed in the mass eigenstates basis both for the charged leptons  $(e, \mu, \tau)$  and for the neutrinos  $(\nu_1, \nu_2, \nu_3)$ so we have the freedom to choose the model. We can choose the basis in which at the beginning $M^l$ or $\mathcal{M}^\nu$ are diagonal. Usually the first alternative is  considered, and we do the same. In the basis in which $M^l$ is diagonal  we can rewrite relation  \eqref{macierzmasowa} as:
\begin{equation}
U^T \mathcal{M}^\nu U = m_{diag}, \label{skrotowej}
\end{equation}
where we introduce the notation:
\begin{equation}
U_{PMNS}=U^\nu\equiv U. 
\end{equation}
The mixing matrix $U$  can be  decomposed in the form:
\begin{equation}
U=f_{NF}\cdot U_D \cdot f_M, \label{macierzdoporownania}
\end{equation} 
where:
\begin{equation}
f_{NF}=\begin{pmatrix}
e^{i \beta_1} & 0 &  0 \\ 
0 & e^{i\beta_2} &  0 \\ 
0 & 0 & e^{i\beta_3}
\end{pmatrix},
\end{equation}
contains the non-physical phases $\beta_i$ which can be absorbed by the charged fermion Dirac fields, and: 
\begin{equation}
f_M=\begin{pmatrix}
1 & 0 & 0  \\ 
0 & e^{(i \alpha_1)/2} & 0  \\ 
0 & 0 & e^{(i \alpha_2)/2}  
\end{pmatrix}
\end{equation}
 includes the Majorana phases $\alpha_i$. Matrix $U_D$  is constructed of  3 mixing angles $\theta_{ij}$ and one Dirac phase $\delta$: 
	\begin{equation}\label{U_D}
	 U_D=\left(
	\begin{matrix}
	c_{12}c_{13} & s_{12}c_{13} & s_{13}e^{-i\delta}\\
	-s_{12}c_{23}-c_{12}s_{13}s_{23}e^{i\delta} & c_{12}c_{23}-s_{12}s_{13}s_{23}e^{i\delta} & c_{13}s_{23}\\
	s_{12}s_{23}-c_{12}s_{13}c_{23}e^{i\delta} & -c_{12}s_{23}-s_{12}s_{13}c_{23}e^{i\delta} & c_{13}c_{23}
	\end{matrix}
	\right).
	\end{equation}
	In such a way, taking into account three neutrino masses, altogether we have 6 real parameters (3 neutrino masses and 3 mixing angles) and 6 phases, exactly equal to the number of free parameters in the  $3\times3$ complex symmetric neutrino mass matrix.
\section{Horizontal Symmetry}
The Standard  Model is up to the  current experimental energies a good working theory. On the other hand it is commonly believed to be only  an effective theory. It agreement with the experimental data  is at the expense of  a huge set of free parameters. We can consider models assuming that both in the lepton and quark sectors there exist fundamental symmetries linking together fermions from different generations giving relations among masses and mixing matrix elements within one family.   If such a symmetry  exists it could reduce the  number of free SM  parameters and shed  light how to extend it. 

Searching of global, primary symmetry (so called horizontal or family symmetry) can be restricted to the lepton sector and actually to testing of the  $U_{PMNS}$ in this sector. As the result of symmetry breaking,  "up" and "down" quarks as well as charged lepton masses, differ significantly among generations. Thus, it is hard to see the symmetry realization. Second of all we could search for it via the CKM matrix but from the experiment we know that off diagonal elements are likely to be small and can be explained as a perturbative corrections beyond SM. 

Generally, from the group point of view, we can distinguish two types of horizontal symmetries: these connected with  finite  discrete abelian groups and these connected with non abelian discrete ones. 

One can take the hypothesis that the shape of the $U_{PMNS}$   and hence the $\mathcal{M}^\nu$ are not accidental. One of the most clear example of the horizontal symmetry realization is related to the TBM matrix story.  Before 2012  oscillation experimental  data were in good agreement with  the known in literature as  $U_{TBM}$ matrix (tribimaximal). It was  proposed in 2002 by  Harison, Perkins and Scott \cite{Harrison:2002er}: 
\begin{equation}
U_{PMNS}\equiv U_{TBM}=\begin{pmatrix}
 \frac{2}{\sqrt{6}} & \frac{1}{\sqrt{3}} & 0 \\
  -\frac{1}{\sqrt{6}} & \frac{1}{\sqrt{3}} & \frac{1}{\sqrt{2}} \\
  \frac{1}{\sqrt{6}} & -\frac{1}{\sqrt{3}} & \frac{1}{\sqrt{2}} 
\end{pmatrix}. \label{TBM}
\end{equation}
The third column corresponds to the maximal mixing between  $\vert \nu_\mu \rangle$ and $\vert \nu_\tau \rangle$ states:
\begin{equation}
\vert U_{\mu 3 } \vert =  \vert U_{\tau 3 } \vert = 1/ \sqrt{2},
\end{equation}
with the mixing angle:
\begin{equation}
\sin \theta_{23}\approx 1/\sqrt{2} \Rightarrow \theta_{23} \approx \pi /4 \textrm{ i.e. }  45^\circ. 
\end{equation}
The second column accounts for an equal mixing of  $\vert \nu_e \rangle$, $\vert \nu_\mu \rangle$ and $\vert \nu_\tau \rangle$ states:
\begin{equation}
\vert U_{e 2 } \vert =  \vert U_{\mu 2} \vert =  \vert U_{\tau 2} \vert =  1/ \sqrt{3} \ ,
\end{equation}
with  the angle $\theta_{12}\approx 35^\circ$.\\
It is worth to mention that:
\begin{equation}
 U_{e 3 }  = 0,
\end{equation}
 so  $\theta_{13}=0$, and   $U_{PMNS}$ is real. 

The shape of the  matrix \eqref{TBM} may suggest the  existence of connection to Clebsch-Gordan coefficients of some symmetry group.   This observation triggered a wide response among the  theoretical groups. There appear lots of phenomenological proposals explaining the TBM pattern. One of the first, put forward in 2004, was the hypothesis linking  $U_{TBM}$ with the  discrete  symmetry group $A_4$  \cite{Ma:2004zv}. It's full description was published in 2005 \cite{Altarelli:2005yp, Altarelli:2005yx, Babu:2005se, deMedeirosVarzielas:2005qg}. \\
 Within almost a decade there have been many other proposals trying better or worse to connect   the $U_{TBM}$ with different symmetry groups: $U(1)^3\times Z^3_2 \rtimes S_3$ \cite{Grimus:2008vg} , $Z_2^3 \rtimes S_3$ \cite{Mohapatra:2006pu}, $S_4$ \cite{Lam:2008rs, Bazzocchi:2008ej}, $T^\prime$ \cite{Feruglio:2007uu, Carr:2007qw}.
In 2012 new evidence for $\theta_{13} \neq 0$  coming from  more  precise oscillation experiments   appear (T2K \cite{Abe:2013hdq},  Daya Bay \cite{An:2013zwz}, MINOS\cite{Adamson:2013whj}   and RENO \cite{Ahn:2012nd}). The angle $\theta_{13}$ was no longer treated as zero.  In the light of new data it turned out that the TBM pattern is completely  ruled out. The question if  it is  possible to find any other symmetry responsible for the lepton  mixing matrix  once more become open.  
   
Up to now, under current experimental data, there are still many  attempts to find valid non abelian horizontal symmetry together for the SM  and it's extensions (see ex.\cite{Dziewit:2013aka}).
 
The models, with the so called  Texture Zeros (TZ),  come back to favor \cite{Nishiura:1999yt, Xing:2002ta, Xing:2002ap}. Through  the mass matrix TZ such patterns are meant   in which   some elements of $\mathcal{M}^\nu$ become zero or almost zero. An universal description on how to obtain the  realization of the symmetry   together for quark and lepton sectors using the TZ can be found in \cite{Grimus:2005sm}.   TZ models are linked with  abelian symmetries  \cite{Grimus:2004hf}. They were always somehow present on the market but because of their triviality (comparing to non abelian symmetries) were less popular during the TBM era. 

Among the many interesting proposals for models with TZ it is worth to mention models assuming vanishing of minors \cite{Lavoura:2004tu}, or the so called hybrid textures  \cite{Grimus:2012zm}. 

From the  methodological point of view  the problem of reconstruction of $\mathcal{M}^\nu$ can be divided into two categories.  In the "top-down" approach the shape of $\mathcal{M}^\nu$
is an assumption emerging from the  proposed theoretical model. All parameters i.e: mixing angles, neutrino masses and phases arise from these assumptions. In contrast  in the "bottom-up" approach the  mass matrix (or the mixing matrix) is built up from the experimental data.  
\section{Bottom-up method}
This section contains some simple example of the "bottom-up" method, in the specific model for neutrino mass. Results presented here are based on \cite{Dziewit:2011pd} and recent experimental data \cite{Tortola:2012te, Fogli:2012ua, GonzalezGarcia:2012sz}. Let us cast the mass matrix in the form: 
\begin{equation}
 \mathcal{M}_\nu=
\left(
\begin{array}{ccc}
 M_{11}e^{i\phi_1} &  M_{12}e^{i\phi_2}& M_{13}e^{i\phi_3} \\
M_{12}e^{i\phi_2} &  M_{22}e^{i\phi_4}& M_{23}e^{i\phi_5} \\
M_{13}e^{i\phi_3} &  M_{23}e^{i\phi_5}& M_{33}e^{i\phi_6}
\end{array}
\right), \label{takam}
\end{equation}
which explicitly depend on six moduli and six phases. Taking into account the reverse relation to equation  \eqref{skrotowej}:
\begin{equation}
\mathcal{M}_\nu= U^\dag m_{diag} U^{-1}, 
\end{equation}
we can express all of them as a function:
\begin{equation}
\left( \mathcal{M}_\nu \right)_{ik} = f_{ik} \left( \theta_{12}, \theta_{13}, \theta_{23}, m_1, m_2, m_3, \delta, \alpha_1, \alpha_2, \beta_1, \beta_2, \beta_3 \right). 
\end{equation}
In this way we get correspondence between all 12 elements of $U$ and $\mathcal{M}^\nu$. Function $f_{ik}$ directly depends on the  Dirac phase $\delta$ as well as Majorana phases $\alpha_1, \alpha_2$ and three non physical phases $\beta_1, \beta_2, \beta_3$. As it was mentioned before we have some information about $\delta$ from the experimental data.  However, there is no  information about the Majorana phases. For huge statistics we can numerically find minima and maxima of  $f_{ik}$  using the following procedure:
\begin{enumerate}
\item
For the given neutrino hierarchy mass, we can express heavier neutrino masses through the lightest one and through known from experimental data mass squared differences. For the normal hierarchy the lightest neutrino mass is $m_1$, so:
\begin{equation}
m_2=\sqrt{m_1^2+\Delta m^2_{21}}, 
\end{equation}
\begin{equation}
m_3=\sqrt{m_1^2+ \Delta m_{31}^2}.
\end{equation}
While for the inverted neutrino mass hierarchy, the lightest  mass is $m_3$, therefore:
\begin{equation}
m_2=\sqrt{m_3^2+ \Delta m_{31}^2+ \Delta m_{21}^2}, 
\end{equation}
\begin{equation}
m_1 = \sqrt{m_3^2+ \Delta m_{31}^2}.
\end{equation}
\item Varying the lightest neutrino mass, all other parameters of $f_{ik}$ are randomly generated within their experimental errors.  
\end{enumerate}
From such a procedure we can get  $\min(f_{ik})$ and $\max(f_{ik})$ dependencies.  
It is assumed that the mass $m_1$ varies in the range: $(0.0001 - 1)$ eV. Function   $f_{ik}$ is treated as $0$ when $f_{ik}<10^{-6}$. 
It is possible to  study  the influence of specific  phases on the function $f_{ik}$. Separately several regions can be set:
\begin{enumerate}
\item 
$\delta$:  $(0-2\pi)$; $\alpha_1=0, \alpha_2=0$,
\item  
  $\delta$: $(0-2\pi)$, $\alpha_1,=\pi/4, \alpha_2=\pi/4$, 
\item 
$\delta,\alpha_1, \alpha_2$: $(0-2\pi)$.
\end{enumerate}
For all above regions $\theta_{12}, \theta_{23}, \theta_{13}$ and $\Delta m_{21}^2$, $\Delta m_{31}^2$  were generated within $1\sigma$ and $3\sigma$ errors.   Examples of obtained dependencies are presented for $\mathcal{M}_{11}$  on figure 1 and  for $\mathcal{M}_{33}$ on figure~2.  Solid grey area represents third  region   at the $1 \sigma$ level. Grey shaded area represents the same region  at the $3 \sigma$ level.  
Studying the  whole set of solutions it can be said that at the  $1\sigma$ significance level  two TZ are excluded. These are: $\mathcal{M}_{11}=0$ for the inverted hierarchy, and $\mathcal{M}_{33}=0$ for the normal hierarchy.  
At the $3\sigma$ significance level only one texture zero is excluded: $\mathcal{M}_{11}=0$ for the normal hierarchy. 
\\
Please notice that the first region does not  distinguished between  Dirac and Majorana neutrinos. For Majorana neutrinos the Dirac phase may be non zero. Non physical phases are not relevant for $|\mathcal{M}_{ik}|$, because elements  $\left( \mathcal{M}_{ik} \right)$ depend on them like:
\begin{equation}
\left( \mathcal{M}_{ik} \right) \sim e^{-i\beta_i} e^{-i \beta_k} \left( \mathcal{M}_{ik} \right)^\prime, 
\end{equation}
where from equation  \eqref{macierzmasowa} and \eqref{macierzdoporownania} we get:
\begin{equation}
\left( \mathcal{M}_{ik} \right)^\prime = \left( U^\ast_D f^\ast_M M_{diag} f^\ast_M U^\dag_D \right)_{ik}.
\end{equation}
There are a lot of  publications (see ex.\cite{Merle:2006du, Grimus:2012ii}) in which neutrino mass matrix elements as a function of the lightest neutrino mass are presented. Part of them are considering influence of CP phases (see ex. \cite{Gozdz}). Just to emphasize the importance of Majorana phases we are presenting plots where  it  can be seen that for an arbitrary chosen values of these phases (second region) elements of the $\mathcal{M}_\nu$ may not be zero. 
\\
Phases  ($\phi_1,\ldots, \phi_6$) are not measurable thus they are  not the subject of analysis.
\section{Summary}
For fixed Majorana phases values (second region)  mass matrix moduli, despite the variation of other parameters,  does not have to go to zero.  In other words Majorana nature is not sufficient for zeroing elements of the neutrino mass matrix. Only in the  some range  of Majorana phases moduli of the neutrino mass matrix can be zero.
\\
Despite the fact that $\theta_{13}$ is different from zero, models with texture zeros are still possible, but the amount of such models is limited.
\\
This work has been supported by the Polish Ministry of Science and Higher Education under grant No. UMO-2013/09/B/ST2/03382.

\newpage
%
%
%
%
\newpage
%
\begin{figure}
\includegraphics[width=10 cm]{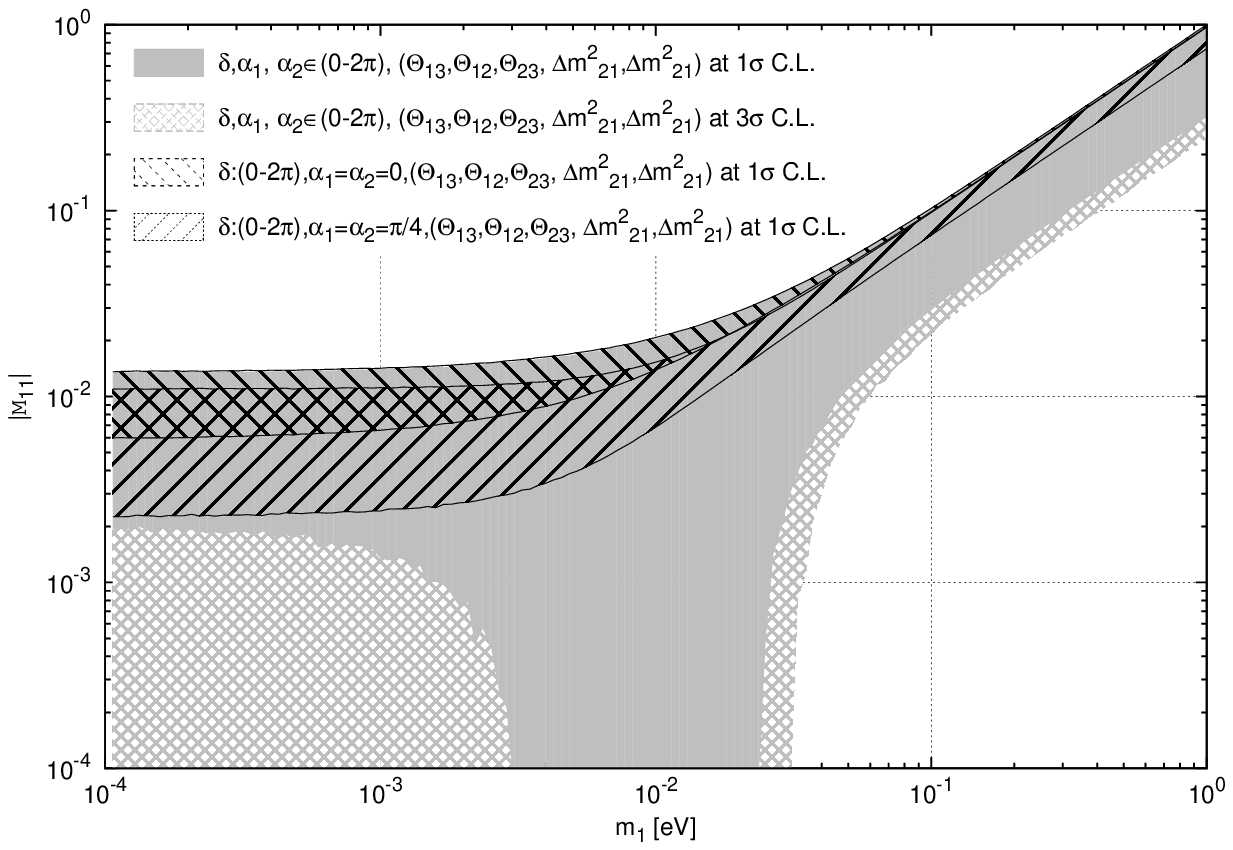}
\caption{Allowed regions of $|\mathcal{M}^\nu_{11}|$ as a function of the lightest  neutrino mass in the case of the normal neutrino mass hierarchy.}
\end{figure}
\newpage
%
\begin{figure}
\includegraphics[width=0.85\textwidth]{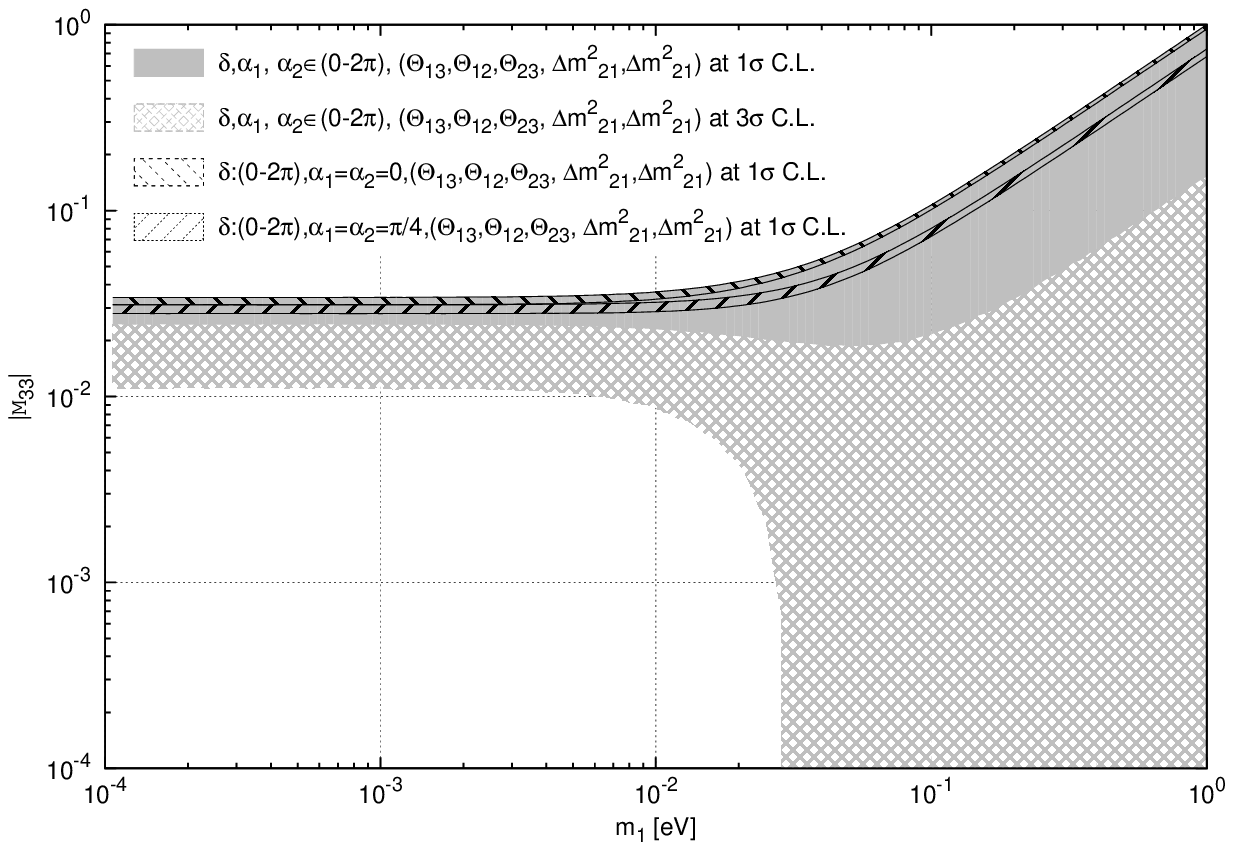}
\caption{Allowed regions of $|\mathcal{M}^\nu_{33}|$ as a function of the lightest  neutrino mass in the case of the normal neutrino mass hierarchy.}
\end{figure}
\newpage
%

\newpage
\begin{thebibliography}{99}
\bibitem{Harrison:2002er} 
  P.~F.~Harrison, D.~H.~Perkins and W.~G.~Scott,
  Phys.\ Lett.\ B {\bf 530}, 167 (2002)
\bibitem{Ma:2004zv} 
  E.~Ma,
  Phys.\ Rev.\ D {\bf 70}, 031901 (2004)
\bibitem{Altarelli:2005yp} 
  G.~Altarelli and F.~Feruglio,
  Nucl.\ Phys.\ B {\bf 720}, 64 (2005)
\bibitem{Altarelli:2005yx} 
  G.~Altarelli and F.~Feruglio,
  Nucl.\ Phys.\ B {\bf 741}, 215 (2006)
\bibitem{Babu:2005se} 
  K.~S.~Babu and X.~G.~He,
\bibitem{deMedeirosVarzielas:2005qg} 
  I.~de Medeiros Varzielas, S.~F.~King and G.~G.~Ross,
  Phys.\ Lett.\ B {\bf 644}, 153 (2007)
\bibitem{Grimus:2008vg} 
  W.~Grimus and L.~Lavoura,
  JHEP {\bf 0904}, 013 (2009)
\bibitem{Mohapatra:2006pu} 
  R.~N.~Mohapatra, S.~Nasri and H.~B.~Yu,
  Phys.\ Lett.\ B {\bf 639}, 318 (2006)
\bibitem{Lam:2008rs} 
  C.~S.~Lam,
  Phys.\ Rev.\ Lett.\  {\bf 101}, 121602 (2008)
\bibitem{Bazzocchi:2008ej} 
  F.~Bazzocchi and S.~Morisi,
  Phys.\ Rev.\ D {\bf 80}, 096005 (2009)
\bibitem{Feruglio:2007uu} 
  F.~Feruglio, C.~Hagedorn, Y.~Lin and L.~Merlo,
  Nucl.\ Phys.\ B {\bf 775}, 120 (2007)
  [Nucl.\ Phys.\  {\bf 836}, 127 (2010)]
\bibitem{Carr:2007qw} 
  P.~D.~Carr and P.~H.~Frampton,
  hep-ph/0701034.
\bibitem{Abe:2013hdq} 
  K.~Abe {\it et al.} [T2K Collaboration],
  Phys.\ Rev.\ Lett.\  {\bf 112}, 061802 (2014)
\bibitem{An:2013zwz} 
  F.~P.~An {\it et al.} [Daya Bay Collaboration],
  Phys.\ Rev.\ Lett.\  {\bf 112}, 061801 (2014)
\bibitem{Adamson:2013whj} 
  P.~Adamson {\it et al.} [MINOS Collaboration],
  Phys.\ Rev.\ Lett.\  {\bf 110}, no. 25, 251801 (2013)
\bibitem{Ahn:2012nd} 
  J.~K.~Ahn {\it et al.} [RENO Collaboration],
  Phys.\ Rev.\ Lett.\  {\bf 108}, 191802 (2012)
\bibitem{Dziewit:2013aka} 
 B.~Dziewit, S.~Zajac and M.~Zralek,
  Acta Phys.\ Polon.\ B {\bf 44}, no. 11, 2353 (2013).
\bibitem{Nishiura:1999yt} 
  H.~Nishiura, K.~Matsuda and T.~Fukuyama,
  Phys.\ Rev.\ D {\bf 60}, 013006 (1999)
\bibitem{Xing:2002ta} 
  Z.~z.~Xing,
  Phys.\ Lett.\ B {\bf 530}, 159 (2002)
\bibitem{Xing:2002ap} 
  Z.~z.~Xing,
  Phys.\ Lett.\ B {\bf 539}, 85 (2002)
\bibitem{Grimus:2005sm} 
  W.~Grimus,
  PoS HEP {\bf 2005}, 186 (2006)    
\bibitem{Grimus:2004hf} 
  W.~Grimus, A.~S.~Joshipura, L.~Lavoura and M.~Tanimoto,
  Eur.\ Phys.\ J.\ C {\bf 36}, 227 (2004)
\bibitem{Lavoura:2004tu} 
  L.~Lavoura,
  Phys.\ Lett.\ B {\bf 609}, 317 (2005)
\bibitem{Grimus:2012zm} 
  W.~Grimus and P.~O.~Ludl,
  J.\ Phys.\ G {\bf 40}, 055003 (2013)
\bibitem{Dziewit:2011pd} 
  B.~Dziewit, S.~Zajac and M.~Zralek,
  Acta Phys.\ Polon.\ B {\bf 42}, 2509 (2011)
\bibitem{Tortola:2012te} 
  D.~V.~Forero, M.~Tortola and J.~W.~F.~Valle,
  Phys.\ Rev.\ D {\bf 86}, 073012 (2012)
\bibitem{Fogli:2012ua} 
  G.~L.~Fogli, E.~Lisi, A.~Marrone, D.~Montanino, A.~Palazzo and A.~M.~Rotunno,
  Phys.\ Rev.\ D {\bf 86}, 013012 (2012)
\bibitem{GonzalezGarcia:2012sz} 
  M.~C.~Gonzalez-Garcia, M.~Maltoni, J.~Salvado and T.~Schwetz,
  JHEP {\bf 1212}, 123 (2012) 
\bibitem{Merle:2006du} 
  A.~Merle and W.~Rodejohann,
  Phys.\ Rev.\ D {\bf 73}, 073012 (2006)
\bibitem{Grimus:2012ii} 
  W.~Grimus and P.~O.~Ludl,
  JHEP {\bf 1212}, 117 (2012)
\bibitem{Gozdz}
  M.~Gozdz, W.~A.~Kaminski and F.~Simkovic,
  Acta Phys.\ Polon.\ B {\bf 37} (2006) 2203
\end{thebibliography}
\end{document}